# A General Sequential Delay-Doppler Estimation Scheme for Sub-Nyquist Pulse-Doppler Radar

Shengyao Chen, Feng Xi, Zhong Liu[*]

Department of Electronic Engineering, Nanjing University of Science and Technology, China

**Abstract**─Sequential estimation of the delay and Doppler parameters for sub-Nyquist radars by analog-to-information conversion (AIC) systems has received wide attention recently. However, the estimation methods reported are AIC-dependent and have poor performance for off-grid targets. This paper develops a general estimation scheme in the sense that it is applicable to all AICs regardless whether the targets are on or off the grids. The proposed scheme estimates the delay and Doppler parameters sequentially, in which the delay estimation is formulated into a beamspace direction-of-arrival problem and the Doppler estimation is translated into a line spectrum estimation problem. Then the well-known spatial and temporal spectrum estimation techniques are used to provide efficient and high-resolution estimates of the delay and Doppler parameters. In addition, sufficient conditions on the AIC to guarantee the successful estimation of off-grid targets are provided, while the existing conditions are mostly related to the on-grid targets. Theoretical analyses and numerical experiments show the effectiveness and the correctness of the proposed scheme.

## I. INTRODUCTION

With the development of analog-to-information conversion (AIC) theory [1], sub-Nyquist radars have acquired wide attention in recent years. Different AIC schemes, such as Xampling [2], random demodulation [3], random modulation pre-integrator (RMPI) [4] and quadrature compressive sampling (QuadCS) [5], have been developed for sampling wideband radar echoes at sub-Nyquist rates. However, the sub-Nyquist samples obtained from AICs are different from the Nyquist samples and, therefore, the traditional processing methods cannot be directly applied for the sub-Nyquist radar processing. For pulse-Doppler radars, several methods for estimating the time delays and Doppler shifts of the targets have been reported [6-11]. In general, these methods can be classified into simultaneous estimation methods [6-8] or sequential methods [9-11]. For simultaneous estimation, the problem is a

---

high-dimensional optimization which requires large computational load. In contrast, the sequential processing performs the estimation of the delay and Doppler parameters separately with a significantly reduced computational load. It has been shown that the sequential methods are more suitable to practical radar applications [10].

It is noticed that each sub-Nyquist sampling method is AIC-oriented and different AICs yield different data structures. Therefore, the sequential methods reported are often AIC-dependent. In [9], a compressive sampling pulse-Doppler (CoSaPD) method is developed for the sub-Nyquist radar with QuadCS. With the assumption that the targets are on the predefined delay grids, CoSaPD first estimates the Doppler shifts using classical spectral analysis techniques and then estimates the time delays through sparse reconstruction. A similar approach, referred to as Doppler focusing (DF), is introduced in [10] for the sub-Nyquist radars with Xampling. Different from CoSaPD, DF assumes a finer grid space. For off-grid targets, both CoSaPD and DF suffer from basis mismatch [12] and their performance deteriorates severely. A powerful sequential method is developed in [11] for the Xampling system. By parametrizing the radar echoes, the method not only efficiently identifies high-resolution time delays and Doppler shifts, but also well solves the off-grid problem. In essence, it formulates the estimation of the time delays and Doppler shifts respectively as a direction-of-arrival (DOA) estimation problem and a line spectral estimation problem. Then the well-developed spatial/temporal spectrum estimation techniques are used to find the time delays/Doppler shifts. In particular, the paper develops a sufficient condition for a successful estimation. However, the theory developed in [11] is based on the special structure of Xampling and is not applicable to other AIC systems [3-5].

Motivated by [11], we present in this paper a general sequential scheme to identify the time delays and Doppler shifts of the radar targets. In addition to the advantages provided by [11], the new scheme is versatile in suiting for any AIC system. The method described in [11] becomes a special case of our scheme. Toward this end, we first define a general compressive parametric model to describe the radar transceiver employing any AIC system. The model incorporates the time delays and Doppler shifts as parameters to be estimated, and groups the radar targets into classes with each corresponding to a distinct time delay. By exploiting the inherent structure of the model, we find that the estimation of the time-delays can be formulated as a beamspace DOA estimation problem [13]. After the time delays are estimated, the estimation of Doppler shifts can be further formulated into a line spectral estimation problem. These formulations resemble those developed in [11]. The difference is that, unlike [11] in which element-space DOA estimation techniques are used for delay estimation, the proposed scheme uses beamspace DOA estimation techniques for this purpose. It is the beamspace DOA formulation that enables us to establish a general estimation scheme. The effectiveness and performance advantages of the proposed scheme over other related methods will be evidently demonstrated through numerical simulations.

Another key contribution of this paper is the derivation of the sufficient condition for the identifiability of the delay parameters. By using the formulation of the beamspace DOA estimation, we theoretically prove that, if the measurement matrix generated by the AIC system satisfies the concentration of measure (CoM) inequality [14] and the number of the compressive measurements is not less than $O(K_\tau \log(N/K_\tau))$, where $K_\tau$ is the number of distinct delays and $N$ is the number of Nyquist samples of radar echo in one pulse repetition interval, the proposed scheme can successfully estimate all time delays. To the best of our knowledge, there are no such results reported in the literature to guarantee the identifiability of the off-grid targets for the general sub-Nyquist pulse-Doppler radar. Several random measurement matrices, such as Gaussian matrix, Bernoulli matrix and sub-Gaussian matrix, are known to satisfy the CoM inequality. The measurement matrices from RMPI and QuadCS also satisfy the CoM inequality as analyzed in Section IV.

The rest of this paper is organized as follows. The problem of interest is formulated in Section II. The proposed estimation scheme is presented in Section III and the sufficient condition for successful estimation is derived in Section IV. Experimental results are simulated in Section V. Finally, conclusions are given in Section VI.

## II. PROBLEM FORMULATION

Consider a co-located pulse-Doppler radar transmitting $L$ pulses in a coherent processing interval (CPI). Assume that there are $K$ non-fluctuating moving point targets in the radar scene. Among the $K$ targets, $K_\tau$ targets have distinct time delays and each time delay $\tau_i$ is associated with $K_{\tau,i}$ different Doppler shifts $v_{ij}$ ( $j = 1, \cdots, K_{\tau,i}$ ). Then the complex baseband echo corresponding to the $l$-th pulse ( $0 \leq l \leq L-1$ ) is given by

$$\tilde{r}^l(t) = \sum_{i=1}^{K_\tau} \sum_{j=1}^{K_{\tau,i}} \alpha_{ij} e^{j2\pi v_{ij} t} \tilde{g}(t - lT - \tau_i) \approx \sum_{i=1}^{K_\tau} \sum_{j=1}^{K_{\tau,i}} \alpha_{ij} e^{j2\pi v_{ij} lT} \tilde{g}(t - lT - \tau_i), \quad t \in [lT, (l+1)T) \quad (1)$$

where $T$ is the pulse repetition interval, $\tilde{g}(t)$ is the complex envelop of transmitting pulses with bandwidth $B$ and duration $T_p$ ( $T_p < T$ ), $\alpha_{ij} \in \mathbb{C}$ denotes the complex reflectivity of the target with the delay-Doppler pair $(\tau_i, v_{ij})$. The approximation follows from the stop-and-hop assumption $v_{ij}T \ll 1$ [15], which implies that $e^{j2\pi v_{ij} t} \approx e^{j2\pi v_{ij} lT}$ for all $t \in [lT,(l+1)T)$. The time delays and Doppler shifts are assumed to be unambiguous, namely, $\tau_i \in [0, T - T_p)$ and $v_{ij} \in (-1/2T, 1/2T)$. The delay resolution and the Doppler resolution are $\tau_0 = 1/B$ and $v_0 = 1/(LT)$, respectively.

Under parametric waveform-matched dictionary [5] with unknown time delay $\tau_i$, $\{\tilde{\psi}(\tau_i,t) | \tilde{\psi}(\tau_i,t) = \tilde{g}(t - \tau_i), i = 1, \cdots, K_\tau\}$, the echo signal $\tilde{r}^l(t)$ can be expressed as

$$\tilde{r}^l(t) = \sum_{i=1}^{K_\tau} \sum_{j=1}^{K_{\tau,i}} \alpha_{ij} e^{j2\pi v_{ij} lT} \tilde{\psi}(\tau_i, t - lT) = \sum_{i=1}^{K_\tau} \alpha_i[l] \tilde{\psi}(\tau_i, t - lT), \quad t - lT \in [0, T) \quad (2)$$

where the sequences $\{\alpha_i[l]\}$, $i = 1, \cdots, K_\tau$, are defined as

$$\alpha_i[l] = \sum_{j=1}^{K_{\tau,i}} \alpha_{ij} e^{j2\pi v_{ij} lT}, \quad l = 0, \ldots, L-1 \quad (3)$$

When the received signal is contaminated by an additive white Gaussian noise (AWGN), the echo $\tilde{r}^l(t)$ is translated into[1]

$$\tilde{r}_1^l(t) = \sum_{i=1}^{K_\tau} \alpha_i[l]\tilde{\psi}(\tau_i, t - lT) + \tilde{n}^l(t), \qquad t - lT \in [0, T) \qquad (4)$$

where $\tilde{n}^l(t)$ is a lowpass complex Gaussian process with power spectral density $N_0$ and bandwidth $B$, and all of the $\tilde{n}^l(t)$ are independently and identically distributed (i.i.d.).

In the framework of sub-Nyquist radars, the echo signal is sampled by an AIC system to attain compressive measurements [16]. Denote $\mathbf{r}^l = \left[\tilde{r}^l(lT), \tilde{r}^l(lT + T_{nyq}), \cdots, \tilde{r}^l(lT + (N-1)T_{nyq})\right]^{\mathrm{T}} \in \mathbb{C}^{N \times 1}$ and $\mathbf{n}^l = \left[\tilde{n}^l(lT), \tilde{n}^l(lT + T_{nyq}), \cdots, \tilde{n}^l(lT + (N-1)T_{nyq})\right]^{\mathrm{T}} \in \mathbb{C}^{N \times 1}$ as the Nyquist-rate sampling vectors of $\tilde{r}^l(t)$ and $\tilde{n}^l(t)$ in the time window $t \in [lT, (l+1)T)$, respectively, where the superscript $(\cdot)^{\mathrm{T}}$ represents transposition. The compressive measurement vector $\mathbf{s}_{cs}^l \in \mathbb{C}^{M \times 1}$ of the echo signal $\tilde{r}_1^l(t)$ obtained from any AIC can be expressed as

$$\mathbf{s}_{cs}^l = \mathbf{M}(\mathbf{r}^l + \mathbf{n}^l) = \mathbf{M}\mathbf{r}^l + \mathbf{n}_{cs}^l \qquad (5)$$

where $\mathbf{M} \in \mathbb{C}^{M \times N}$ ($M < N$) is the measurement matrix depending on the employed AIC system [2-5] and $\mathbf{n}_{cs}^l = \mathbf{M}\mathbf{n}^l = \left[\tilde{n}_{cs}^l[1], \tilde{n}_{cs}^l[2], \cdots, \tilde{n}_{cs}^l[M]\right]^{\mathrm{T}} \in \mathbb{C}^{M \times 1}$ is the compressive measurement vector of the noise $\tilde{n}^l(t)$. Note that $\tilde{n}_{cs}^l[m]$ ($1 \leq m \leq M$) is an i.i.d. complex Gaussian process with zero-mean and variance $NN_0B/M$ [17].

Denote $\boldsymbol{\psi}(\tau_i) = \left[\tilde{\psi}(\tau_i, 0), \tilde{\psi}(\tau_i, T_{nyq}), \cdots, \tilde{\psi}(\tau_i, (N-1)T_{nyq})\right]^{\mathrm{T}} \in \mathbb{C}^{N \times 1}$ as the Nyquist-rate sampling vector of the parametric atom $\tilde{\psi}(\tau_i, t)$. With (2), we have $\mathbf{r}^l = \sum_{i=1}^{K_\tau} \alpha_i[l]\boldsymbol{\psi}(\tau_i)$. Then the compressive vector (5) can be explicitly described by

$$\mathbf{s}_{cs}^l = \sum_{i=1}^{K_\tau} \alpha_i[l]\mathbf{M}\boldsymbol{\psi}(\tau_i) + \mathbf{n}_{cs}^l \qquad (6)$$

---

[1] In practical scenario, the received signal may also contain clutter in addition to noise. We will discuss the effect of the clutter in Section V-D.

Let $\mathbf{\Psi} = [\boldsymbol{\psi}(\tau_1), \boldsymbol{\psi}(\tau_2), \cdots, \boldsymbol{\psi}(\tau_{K_\tau})] \in \mathbb{C}^{N \times K_\tau}$ and $\boldsymbol{\theta}^l = [\alpha_1[l], \alpha_2[l], \cdots, \alpha_{K_\tau}[l]]^T \in \mathbb{C}^{K_\tau \times 1}$. We can rewrite (6) as

$$\mathbf{s}_{cs}^l = \mathbf{M}\mathbf{\Psi}\boldsymbol{\theta}^l + \mathbf{n}_{cs}^l \tag{7}$$

Concatenating all $L$ compressive measurement vectors $\mathbf{s}_{cs}^l$ and $\mathbf{n}_{cs}^l$ as $\mathbf{S} = [\mathbf{s}_{cs}^0, \mathbf{s}_{cs}^1, \cdots, \mathbf{s}_{cs}^{L-1}] \in \mathbb{C}^{M \times L}$ and $\mathbf{N} = [\mathbf{n}_{cs}^0, \mathbf{n}_{cs}^1, \cdots, \mathbf{n}_{cs}^{L-1}] \in \mathbb{C}^{M \times L}$, respectively, we have the compressive measurement of the radar echo in a CPI as

$$\mathbf{S} = \mathbf{M}\mathbf{\Psi}\boldsymbol{\Theta} + \mathbf{N} \tag{8}$$

with $\boldsymbol{\Theta} = [\boldsymbol{\theta}^0, \boldsymbol{\theta}^1, \cdots, \boldsymbol{\theta}^{L-1}] \in \mathbb{C}^{K_v \times L}$.

Our goal is to accurately estimate the unknown delay-Doppler pairs $(\tau_i, v_{ij})$ and the corresponding complex reflectivity $\alpha_{ij}$ from the compressive measurement data matrix $\mathbf{S}$. Let $\mathbf{s}' = \text{vec}(\mathbf{S}) \in \mathbb{C}^{ML \times 1}$ and $\mathbf{n}' = \text{vec}(\mathbf{N}) \in \mathbb{C}^{ML \times 1}$ be the column-wise vectorized $\mathbf{S}$ and $\mathbf{N}$, respectively. Based on (3) and (6), we can express $\mathbf{s}'$ as

$$\mathbf{s}' = \sum_{i=1}^{K_\tau} \sum_{j=1}^{K_{\tau,i}} \alpha_{ij} \mathbf{b}(v_{ij}) \otimes \mathbf{M}\boldsymbol{\psi}(\tau_i) + \mathbf{n}' \tag{9}$$

where $\mathbf{b}(v_{ij}) = [1, e^{j2\pi v_{ij}T}, \cdots, e^{j2\pi v_{ij}(L-1)T}]^T$ and $\otimes$ denotes the Kronecker product. Then the delay-Doppler pairs $(\tau_i, v_{ij})$ and reflectivities $\alpha_{ij}$ can be found by solving the following high-dimensional optimization problem

$$\min_{\tau_i, v_{ij}, \alpha_{ij}} \left\| \mathbf{s}' - \sum_{i=1}^{K_\tau} \sum_{j=1}^{K_{\tau,i}} \alpha_{ij} \mathbf{b}(v_{ij}) \otimes \mathbf{M}\boldsymbol{\psi}(\tau_i) \right\|_2^2 \tag{10}$$

using the compressive parameter estimation techniques [8, 18, 19]. In addition to a high computational load, however, a big challenge in applying these techniques is that there are no theoretical guarantees of the estimated off-grid parameters. To solve this problem, in the following, we will develop an efficient scheme to sequentially estimate time-delays, Doppler shifts and reflectivities. Sufficient conditions for successful estimation are also presented.

## III. SEQUENTIAL ESTIMATION SCHEME

The proposed scheme first estimates the time delays, then the Doppler shifts and finally the reflectivities. However, as noted from (9), once the delay and Doppler parameters are determined, the estimation of the reflectivities reduces to a simple least squares fit which is well documented in the literature (see, for example, [20]). Therefore, in the following, we mainly address the estimation of the delay-Doppler pairs. In addition, we also discuss the realization of the proposed scheme.

### A. Estimation of Time Delays with Beamspace DOA Estimation

The keystone of the proposed scheme is to identify the time delays by solving a beamspace DOA estimation problem.

Let $\tilde{g}[i] = \tilde{g}(iT_{nyq})$ and $\tilde{g}'[i] = \tilde{g}(iT_{nyq} - \tau)$ ($i = 0,1,\cdots,N-1$) be the Nyquist samples of the transmitted pulse $\tilde{g}(t)$ and its time-delay version $\tilde{g}(t-\tau)$, respectively. Denote $G[k]$ and $G'[k]$ as the discrete Fourier transform (DFT) of the sequence $\tilde{g}[i]$ and $\tilde{g}'[i]$ ($i = 0,1,\cdots,N-1$). It is clear that $G'[k] = G[k]e^{-j2\pi k\tau/(NT_{nyq})} = G[k]e^{-j2\pi k\tau/T}$. Define $\mathbf{G} = diag(G[0], G[1], \cdots, G[N-1])$. The DFT $\hat{\boldsymbol{\psi}}(\tau_i)$ of the Nyquist sample vector $\boldsymbol{\psi}(\tau_i)$ of $\tilde{\psi}(\tau_i, t)$ or $\tilde{g}(t-\tau_i)$ is given by

$$\hat{\boldsymbol{\psi}}(\tau_i) = \mathbf{G}\mathbf{a}(\tau_i) \tag{11}$$

where $\mathbf{a}(\tau_i) = \left[1, e^{-j2\pi\tau_i/T}, \cdots, e^{-j2\pi(N-1)\tau_i/T}\right]^T$. Denote $\mathbf{A} = \left[\mathbf{a}(\tau_1), \mathbf{a}(\tau_2), \cdots, \mathbf{a}(\tau_{K_\tau})\right] \in \mathbb{C}^{N \times K_\tau}$. The matrix $\boldsymbol{\Psi}$ in (8) can be decomposed as

$$\boldsymbol{\Psi} = \mathbf{F}^{-1}\mathbf{G}\mathbf{A} \tag{12}$$

where $\mathbf{F}^{-1} \in \mathbb{C}^{N \times N}$ is the inverse DFT matrix with the $(i,k)$-th element being $\mathbf{F}_{ik}^{-1} = 1/N \, e^{j2\pi(i-1)(k-1)/N}$. Substituting (12) into (8), we obtain

$$\mathbf{S} = \mathbf{M}\mathbf{F}^{-1}\mathbf{G}\mathbf{A}\boldsymbol{\Theta} + \mathbf{N} \tag{13}$$

If we take $\mathbf{A}$ as the steering matrix, $\boldsymbol{\Theta}$ as the source matrix consisting of $K_\tau$ sources and $\mathbf{M}\mathbf{F}^{-1}\mathbf{G}$ as the beamformer matrix, we can see that (13) is exactly the same as the model of the beamspace DOA estimation problem for an *N*-element uniform linear array (ULA). The $i$-th row of $\Theta$ represents the arriving source with spatial frequency

$$f_i = \begin{cases} \tau_i/T, & 0 \leq \tau_i < T/2 \\ \tau_i/T - 1, & T/2 \leq \tau_i < T \end{cases}$$

Suppose that $\hat{f}_i$ is the estimated spatial frequency of the $i$-th source. Then the corresponding delay estimate is given by $\hat{\tau}_i = \text{T} \mod(\hat{f}_i + 1, 1)$, where "mod" is the remainder function.

From the beamspace DOA estimation theory [21], we know that, for a successful estimation, the matrix $\mathbf{MF}^{-1}\mathbf{GA} = \mathbf{M\Psi} \in \mathbb{C}^{M \times K_v}$ must be of full-column rank to ensure the dimension of signal subspace not to decrease after the beamformer transformation. When classical beamspace DOA estimation algorithms[2], such as beamspace MUSIC algorithm [13], are used to estimate the time delays, in addition to this requirement, we further require that the noise $\mathbf{N}$ has the i.i.d. property, and $\Theta$ is of full-row rank, *i.e.*, the arriving sources are incoherent [13]. While the i.i.d. noise requirement is satisfied [17], the property of full-row rank $\Theta$ needs to be confirmed. The sufficient conditions to ensure the estimation of the delay parameters are derived in Section VI.

The estimation method in [11] is a special case of the proposed scheme for the Xampling system. For Xampling AIC, its measurement matrix $\mathbf{M}$ is a partial Fourier matrix consisting of the first consecutive $M$ rows of the DFT matrix. Inserting the matrix $\mathbf{M}$ into (13), we have $\mathbf{MF}^{-1} = [\mathbf{I}_M, \mathbf{O}]$, where $\mathbf{I}_M \in \mathbb{C}^{M \times M}$ is an identity matrix and $\mathbf{O} \in \mathbb{C}^{M \times (N-M)}$ is a zero matrix. Then we can derive $\mathbf{M\Psi} = \mathbf{G}_M \mathbf{A}_M$, where $\mathbf{G}_M = diag(G[0], G[1], \cdots, G[M-1])$ and $\mathbf{A}_M$ is a Vandermonde matrix consisting of the first $M$ rows of $\mathbf{A}$. In other words, for Xampling AIC, the beamspace DOA formulation (13) with $N$ array elements degenerates into the element-space DOA problem with $M$ array elements.

---

[2] In DOA estimation problems, the simplest case is that the noise is i.i.d. and the arriving sources are incoherent. For convenience, we call the DOA estimation algorithms for such case as the classical algorithms.

## B. Estimation of Doppler Shifts with Line Spectrum Estimation

After finding the time delays, we can now extract the noisy version of $\Theta$ from the data matrix $\mathbf{S}$ in (8) through the following relationship

$$\hat{\Theta} = (\mathbf{M}\Psi)^\dagger \mathbf{S} = \Theta + (\mathbf{M}\Psi)^\dagger \mathbf{N} \tag{14}$$

where $(\cdot)^\dagger$ is the Moore-Penrose pseudo-inverse, and $(\mathbf{M}\Psi)^\dagger \mathbf{M}\Psi = \mathbf{I}$ is satisfied for the full-column rank matrix $\mathbf{M}\Psi$. Denote $\boldsymbol{\alpha}_i^T = [\alpha_i[0], \alpha_i[1], \cdots, \alpha_i[L-1]]$ as the $i$-th row vector of $\Theta$. Similarly, let $\hat{\boldsymbol{\alpha}}_i^T$ and $\mathbf{n}_i^T$ be the $i$-th row vector of $\hat{\Theta}$ and $(\mathbf{M}\Psi)^\dagger \mathbf{N}$, respectively. For a given $\tau_i$, we have from (3)

$$\hat{\boldsymbol{\alpha}}_i = \boldsymbol{\alpha}_i + \mathbf{n}_i = \sum_{j=1}^{K_{\tau,i}} \alpha_{ij} \mathbf{b}(v_{ij}) + \mathbf{n}_i \tag{15}$$

It is seen that the $i$-th row of $\hat{\Theta}$ is only related to the Doppler shifts $v_{ij}$ ($j = 1, \cdots, K_{\tau,i}$). As a result, we can independently estimate the Doppler shifts $v_{ij}$ in a class corresponding to each time delay $\tau_i$. It is noted that formulation (15) is exactly a parametric line spectral estimation problem. As such, various techniques, such as annihilating-filter method and subspace methods (see [22] for a review), can be used to estimate the Doppler shifts $v_{ij}$ ($j = 1, \cdots, K_{\tau,i}$).

For the solution of (15) to exist, it is known that the condition $L \geq 2\max(K_{\tau,i})$ must hold [11]. The condition is easily satisfied in practical radar systems.

## C. Estimation of Reflectivities with Least-Squares Fit

With the estimated time delays $\hat{\tau}_i$ and Doppler shifts $\hat{v}_{ij}$, the reflectivities $\alpha_{ij}$ can be easily estimated by the least-squares fit as

$$\min_{\alpha_{ij}} \left\| \mathbf{s}' - \sum_{i=1}^{K_\tau} \sum_{j=1}^{K_{\tau,i}} \alpha_{ij} \mathbf{b}(\hat{v}_{ij}) \otimes \mathbf{M}\boldsymbol{\psi}(\hat{\tau}_i) \right\|_2^2 \tag{16}$$

## D. Practical Considerations

We first discuss the realizations of the proposed scheme. For the estimation of the time-delays, we take the beamspace root MUSIC algorithm [23] or the beamspace spectral MUSIC algorithm [13] as realization examples[3]. For the estimation of the Doppler shifts, we utilize the ESPRIT algorithm [25] because of its high estimation accuracy and computational efficiency. The computation of the target reflectivities is just the least-squares fitting. Then we come to two realizations of the proposed scheme: beamspace root MUSIC + ESPRIT and beamspace spectral MUSIC + ESPRIT. For brevity, we denote the first one as GeSeDD-1 and the second one as GeSeDD-2 with GeSeDD representing general sequential delay-Doppler estimation.

Next we study the computational complexity of the two realizations. As we know, the beamspace MUSIC algorithms mainly consist of two calculations: eigen-decomposition and DOA searching. The computational complexity of the eigen-decompostion equals to $\mathcal{O}(L^2 M + M^3)$. The search complexity by the root-finding is $\mathcal{O}(N^3)$ [26] and the complexity by the grid search is $\mathcal{O}(DMN^2)$, where $D$ is the ratio between the Nyquist grid space and the grid-search space. For the estimation of the Doppler shifts, we should first extract the noisy coefficient matrix $\hat{\Theta}$ from $\mathbf{S}$ which will take $\mathcal{O}(KM^2 + KLM)$ operations. In the worst case where the time-delays and Doppler shifts are distinct between any two targets, estimating the Doppler shifts from $\hat{\Theta}$ by the ESPRIT algorithm will take $\mathcal{O}(KL^3)$ operations. For the calculations of the target reflectivities, the cost of least squares fit equals to $\mathcal{O}(K^2 LM)$. Then the GeSeDD-1 method will take $\mathcal{O}(L^2 M + M^3) + \mathcal{O}(N^3) + \mathcal{O}(KM^2 + KLM) + \mathcal{O}(KL^3) + \mathcal{O}(K^2 LM)$ operations. Similarly, the GeSeDD-2 method takes $\mathcal{O}(L^2 M + M^3) + \mathcal{O}(DMN^2) + \mathcal{O}(KM^2 + KLM) + \mathcal{O}(KL^3) + \mathcal{O}(K^2 LM)$ operations. For sub-Nyquist radar, it is known that $K(\text{or } D) < L \ll M < N$. Then the computation complexities of GeSeDD-1 and GeSeDD-2 can be approximated as $\mathcal{O}(N^3)$ and $\mathcal{O}(DMN^2)$, respectively. Along the same lines, we can derive the complexities of CoSaPD and DF, which are respectively about $\mathcal{O}(KQMN)$ and $\mathcal{O}(KD^2 LMN)$,

---

[3] In beamspace DOA estimation, we know that the beamspace ESPRIT is an efficient DOA estimation algorithm [24]. However, for the formulation (13), the algorithm cannot be used to find the time-delays because of its special beamformer matrix.

where $Q$ is the maximum number of iterations for convex optimization sub-problems in CoSaPD and satisfies $L < Q < M$. With the above analysis, we can find that GeSeDD-1 has larger computational complexity than CoSaPD and DF, and GeSeDD-2 is comparable to CoSaPD and DF. In the proposed realizations, GeSeDD-1 and GeSeDD-2 respectively utilize the root finding and the grid search to find the time-delays. Therefore, GeSeDD-1 has large computational complexity. However, GeSeDD-1 has high estimation accuracy because of the root finding, while the estimation accuracy by GeSeDD-2 is limited by the grid space. These observations are validated by simulation results in Section V-C.

Note that in both GeSeDD-1 and GeSeDD-2, we have to know in advance the number $K_\tau$ in the time-delay estimation and the number $K_{\tau,i}$ ($i = 1, \cdots, K_\tau$) in the Doppler estimation. This problem is equivalent to the model order selection in spectrum analysis and can be resolved by exploiting the Akaike information criterion or the Bayesian information criterion [22].

## IV. SUFFICIENT CONDITIONS FOR SUCCESSFUL ESTIMATION

In this section, we provide the sufficient conditions to ensure the full-column rank of $\mathbf{M\Psi}$ and the full-row rank of $\mathbf{\Theta}$.

Before proceeding, we introduce two basic concepts which are widely utilized in the analysis of measurement matrix $\mathbf{M}$. These two concepts are fundamentals to derive the full-column rank property of $\mathbf{M\Psi}$.

**Definition 1** [27]**:** For each integer $k = 1, 2, \cdots$, define the isometry constant $\delta_k$ of the matrix $\mathbf{M}$ as the smallest number such that

$$\left(1 - \delta_K\right)\|\mathbf{x}\|_2^2 \leq \|\mathbf{Mx}\|_2^2 \leq \left(1 + \delta_K\right)\|\mathbf{x}\|_2^2 \tag{17}$$

holds for all $k$-sparse vectors of $\mathbf{x}$. If $\delta_K \in (0,1)$, we say that the matrix $\mathbf{M}$ obeys the restricted isometry property (RIP) of order $K$.

The RIP quantifies how well the matrix $\mathbf{M}$ preserves the norm of sparse vectors $\mathbf{x}$. For the matrix $\mathbf{M}$ satisfying the RIP of order $K$, the sub-matrix $\mathbf{M}_\Lambda$ consisting of arbitrary $K$ columns of $\mathbf{M}$ is necessarily full-column rank because $\|\mathbf{M}_\Lambda \mathbf{x}_\Lambda\|_2^2 / \|\mathbf{x}_\Lambda\|_2^2 \geq 1 - \delta_K > 0$ for any $\mathbf{x}_\Lambda$, where $\Lambda$ is the index set of column vectors.

**Definition 2** [28]**:** For a random matrix $\mathbf{M}$, if the concentration of measure (CoM) inequality

$$\mathrm{P}\left\{\left|\|\mathbf{Mx}\|_2^2 - \|\mathbf{x}\|_2^2\right| \geq \varepsilon \|\mathbf{x}\|_2^2\right\} \leq 2e^{-Mc_0(\varepsilon)}, \quad 0 < \varepsilon < 1 \tag{18}$$

holds for any $\mathbf{x} \in \mathbb{C}^N$, where $c_0(\varepsilon) > 0$ is a constant depending only on $\varepsilon$, we say that the random matrix $\mathbf{M}$ satisfies the CoM inequality.

The CoM inequality is a powerful tool to analyze the RIP of random matrices. Theorem 5.2 in [14] states that $\mathbf{M}$ satisfies the RIP of order $K$ with a high probability if it satisfies the CoM inequality (18) and $M = O(K \log(N/K))$. With these concepts, we can prove the following theorem.

**Theorem 1:** For a given $\delta_{K_\tau} \in (0,1)$, if a random matrix $\mathbf{M}$ satisfies the CoM inequality (18) and $M \geq c_1 K_\tau \log(N/K_\tau)$, then the matrix $\mathbf{M\Psi}$ has a full-column rank with a probability no less than $1 - e^{-c_2 M}$, where $c_1$ and $c_2$ are positive constants depending only on the given $\delta_{K_\tau}$.

***Proof:*** From (12), we know that $\mathbf{F}^{-1}$ and $\mathbf{G}$ are full-rank square matrices, and $\mathbf{A}$ has a full-column rank since it is a Vandermonde matrix. Then the matrix $\mathbf{\Psi} = \mathbf{F}^{-1}\mathbf{GA}$ has a full-column rank. Performing the singular value decomposition on $\mathbf{\Psi}$, we have

$$\mathbf{\Psi} = \mathbf{U}\begin{bmatrix}\mathbf{\Sigma}\\\mathbf{O}\end{bmatrix}\mathbf{V}^H = \mathbf{U}_{K_\tau}\mathbf{\Sigma}\mathbf{V}^H \tag{19}$$

where $\mathbf{U} \in \mathbb{C}^{N \times N}$ and $\mathbf{V} \in \mathbb{C}^{K_\tau \times K_\tau}$ are unitary matrices, $\mathbf{\Sigma} \in \mathbb{C}^{K_\tau \times K_\tau}$ is a diagonal matrix whose diagonal entries are the non-zero singular-values of $\mathbf{\Psi}$, $\mathbf{O} \in \mathbb{C}^{(N-K_\tau) \times K_\tau}$ is a zero matrix, and $\mathbf{U}_{K_\tau}$ is the matrix composed of the first $K_\tau$ columns of $\mathbf{U}$. Using the result of (19), we can obtain

$$\mathbf{M\Psi} = \mathbf{MU}_{K_\tau}\mathbf{\Sigma}\mathbf{V}^H \tag{20}$$

It is clear that $\mathbf{M\Psi}$ has a full-column rank if and only if $\mathbf{MU}_{K_\tau}$ is a full-column rank matrix.

Now, we derive the condition to ensure the full-column rank of $\mathbf{MU}_{K_\tau}$. For the random matrix $\mathbf{M}$

satisfying the CoM inequality (18), we can represent any $\mathbf{x}$ in terms of an orthogonal basis $\mathbf{U}$ as $\mathbf{x} = \mathbf{U}\boldsymbol{\beta}$. Inserting this relationship into (18), we find that the matrix $\mathbf{MU}$ also satisfies the CoM inequality. According to Theorem 5.2 in [14], if $M \geq c_1 K_\tau \log(N/K_\tau)$, $\mathbf{MU}$ satisfies the RIP of order $K_\tau$ with probability no less than $1 - e^{-c_2 M}$, where $c_1$ and $c_2$ are positive constants depending only on the prescribed $\delta_{K_\tau}$. Then the matrix $\mathbf{MU}_{K_\tau}$ is necessarily of full-column rank. The proof is completed. □

Theorem 1 is of fundamental importance. If the measurement matrix $\mathbf{M}$ of an AIC system satisfies the CoM inequality, we can successfully estimate the time delays using the beamspace DOA estimation algorithms with a high probability. It has been shown that the CoM inequality holds for several kinds of random matrices, such as Gaussian matrix, Bernoulli matrix and sub-Gaussian matrix [14]. For typical AIC systems, we have some aggressive results to ensure their measurement matrices to satisfy the CoM inequality. For the RMPI system [4], each channel can produce a compressive measurement value and the measurement matrix is totally random. Then the measurement matrix by RMPI satisfies the CoM inequality. For the QuadCS system, we have revealed that its measurement matrix manifests a random circulant structure in the frequency domain and thus satisfies the CoM inequality in [5].

Hereafter, we deduce the incoherent source condition for the classical beamspace DOA estimation algorithms.

**Theorem 2:** The matrix $\Theta$ is of full-row rank if in a target class corresponding to the time-delay $\tau_i$, there exits at least a Doppler shift distinct from all the other $K-1$ Doppler shifts.

***Proof***: The full-row rank of $\Theta$ is equivalent to linear independence of all the row vectors $\boldsymbol{\alpha}_i^\mathrm{T}$. That is to say that $\sum_{i=1}^{K_\tau} k_i \boldsymbol{\alpha}_i = 0$ only if all the constants $k_i$ $(i=1,2,\cdots,K_\tau)$ are equal to zero.

For the target class corresponding to the time-delay $\tau_i$, without loss of generality, assume that $v_{i1}$

is the distinct Doppler shift. As defined in (15), $\mathbf{\alpha}_i = \sum_{j=1}^{K_{\tau,i}} \alpha_{ij} \mathbf{b}(v_{ij})$. Then $\sum_{i=1}^{K_\tau} k_i \mathbf{\alpha}_i = 0$ becomes

$$\sum_{i=1}^{K_\tau} \alpha_{i1} k_i \mathbf{b}(v_{i1}) + \sum_{i=1}^{K_\tau} \sum_{j=2}^{K_{\tau,i}} k_i \alpha_{ij} \mathbf{b}(v_{ij}) = 0 \tag{21}$$

Note that $\mathbf{b}(v_{ij}) = \left[1, e^{j2\pi v_{ij} T}, \cdots, e^{j2\pi v_{ij}(L-1)T}\right]^\mathrm{T}$. Then for any $i$, $i = 1, 2, \cdots, K_\tau$, $\mathbf{b}(v_{i1})$ is independent of $\mathbf{b}(v_{ij})$ for $i = 1, 2, \cdots, K_\tau$ and $j = 2, 3, \cdots, K_{\tau,i}$ because $v_{i1} \neq v_{ij}$ ($i = 1, 2, \cdots, K_\tau$; $j = 2, 3, \cdots, K_{\tau,i}$). Therefore,

$$\sum_{i=1}^{K_\tau} \alpha_{i1} k_i \mathbf{b}(v_{i1}) = 0 \tag{22}$$

As $\alpha_{i1}$ is non-zero and all elements of $\mathbf{b}(v_{i1})$ ($i = 1, 2, \cdots, K_\tau$) are linearly independent, (22) holds only if $k_i = 0$ ($i = 1, 2, \cdots, K_\tau$). As a result, all the vectors $\mathbf{\alpha}_i$ are linearly independent and, thereby, $\Theta$ is of full-row rank. The proof is completed. $\square$

Theorem 2 is derived from the requirement of the classical beamspace DOA estimation algorithms. In practical applications, for example, in the detection and estimation of multiple radar targets, this requirement is easily satisfied. When the radar targets result in coherent signals, these algorithms fail to estimate the delay parameters. In this case, more advanced techniques should be developed. A popular solution is to use the spatial smoothing technique [29]. However, for the developed model (13), we cannot directly perform spatial smoothing because the element space data $\mathbf{A}\Theta$ in (13) is not available. One possible way is to interpolate the data $\mathbf{S}$ in (13) as done in the interpolated array, i.e., obtaining the interpolated data by formulating a DOA estimation problem for ULA with $M$ virtual array elements [30]. Then the DOA estimation technique using spatial smoothing with the interpolated array [31] can be used to find the estimates of the delay parameters.

## V. SIMULATION RESULTS

In this section, we present several numerical experiments to demonstrate the performance of the proposed scheme. Subsections V-A and V-B respectively provide the estimation and resolution performance of the time-delays and Doppler shifts in the AWGN environments. Subsections V-C tests the computational loads. The effects of clutter are examined in Subsection V-D.

We simulate the performance of the proposed two realizations, GeSeDD-1 and GeSeDD-2, and make a comparison with CoSaPD in [9] and DF in [10]. For GeSeDD-2, the grid space is set as one-fifth delay resolution for grid search in beamspace spectral MUSIC. For the DF method, the grid spaces of time delay and Doppler shift are both set in a half of resolution. As a benchmark, a simultaneous estimation method, the parameter perturbed orthogonal matching pursuit (PPOMP) algorithm in [8], is also shown for its high estimation accuracy.

In simulation experiments, the QuadCS is taken as a sample AIC. The sampling rate is set to be one-fifth of the Nyquist rate. The received signal $\tilde{r}^l(t)$ is assumed to be a linear combination of $K$ delayed and Doppler-shifted versions of the linear frequency modulation (LFM) pulse with bandwidth $B = 100\,\text{MHz}$ and pulse width $T_p = 10\,\mu\text{s}$. The pulse repetition interval is $T = 100\,\mu\text{s}$ and a CPI consists of 100 pulses. With these assumed parameters, the unambiguous time delays and Doppler shifts are obtained as $0 \sim 90\,\mu\text{s}$ and $-5 \sim 5\,\text{KHz}$, respectively. The delay resolution $\tau_0$ is $0.01\,\mu\text{s}$ and the Doppler resolution $v_0$ is $0.1\,\text{KHz}$. Unless otherwise specified, the time delays and the Doppler shifts of all targets are randomly chosen with a uniform distribution from the intervals $(0,10)\,\mu\text{s}$ and $(-5,5)\,\text{KHz}$. The reflectivity amplitudes and phases of the targets are uniformly distributed in $[0.1,1]$ and $(0,2\pi]$, respectively. The received signal-to-noise ratio (SNR) is defined as $\text{SNR}^l = \frac{1}{T}\int_{(l-1)T}^{lT}|\tilde{r}^l(t)|^2 dt \Big/ (N_0 B)$, which is assumed to be fixed in different pulse intervals. The $\text{SNR}^l$ equals to SNR in compressive domain, $\text{SNR}_{cs}^l = \|\mathbf{Mr}^l\|_2^2 \Big/ \|\mathbf{n}_{cs}^l\|_2^2$, as revealed in [5]. We omit its superscript and subscript below.

The relative root-mean-square error (RRMSE) is taken as a metric to evaluate the performance. Let $\bar{\tau}_k$ and $\bar{v}_k$ be the estimate of the time-delay and the Doppler shift of the $k$-th target. The RRMSEs for the time-delays and the Doppler shifts are defined respectively as

$$\text{RRMSE}_\tau = \frac{1}{\tau_0}\sqrt{\frac{1}{K}\sum_{k=1}^{K}(\bar{\tau}_k - \tau_k)^2} \quad \text{and} \quad \text{RRMSE}_v = \frac{1}{v_0}\sqrt{\frac{1}{K}\sum_{k=1}^{K}(\bar{v}_k - v_k)^2} \tag{23}$$

For the scene of $K$ targets, the targets are detected from the $K$ largest amplitudes of reflectivity.

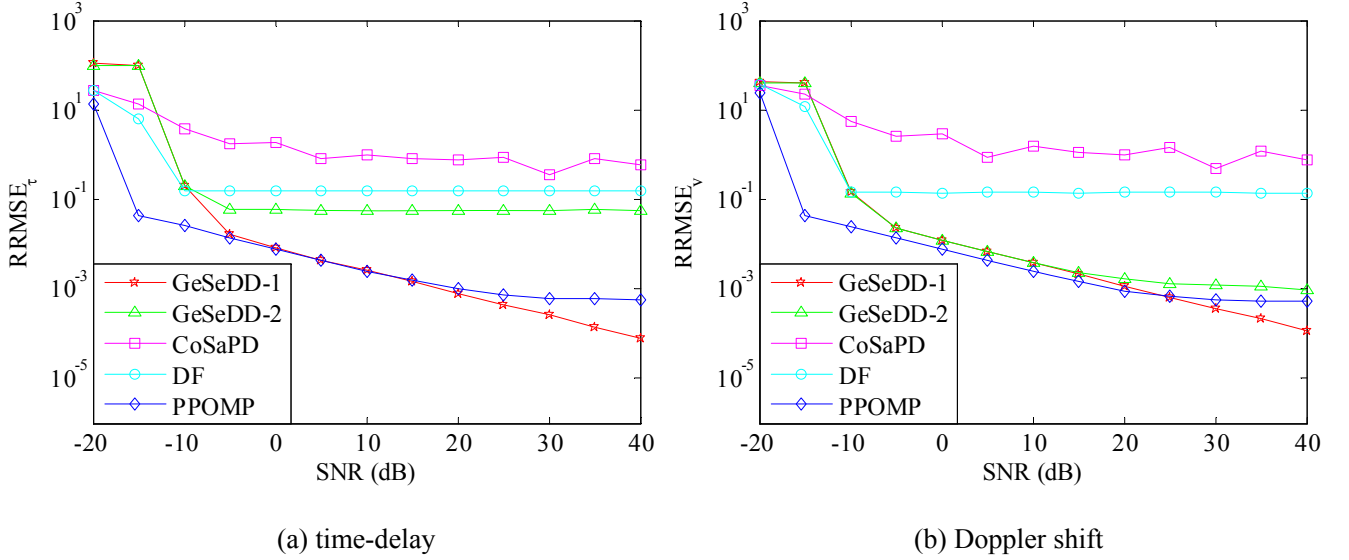

(a) time-delay       (b) Doppler shift

Fig. 1 The RRMSE versus the SNR in a noisy case ($K$=10)

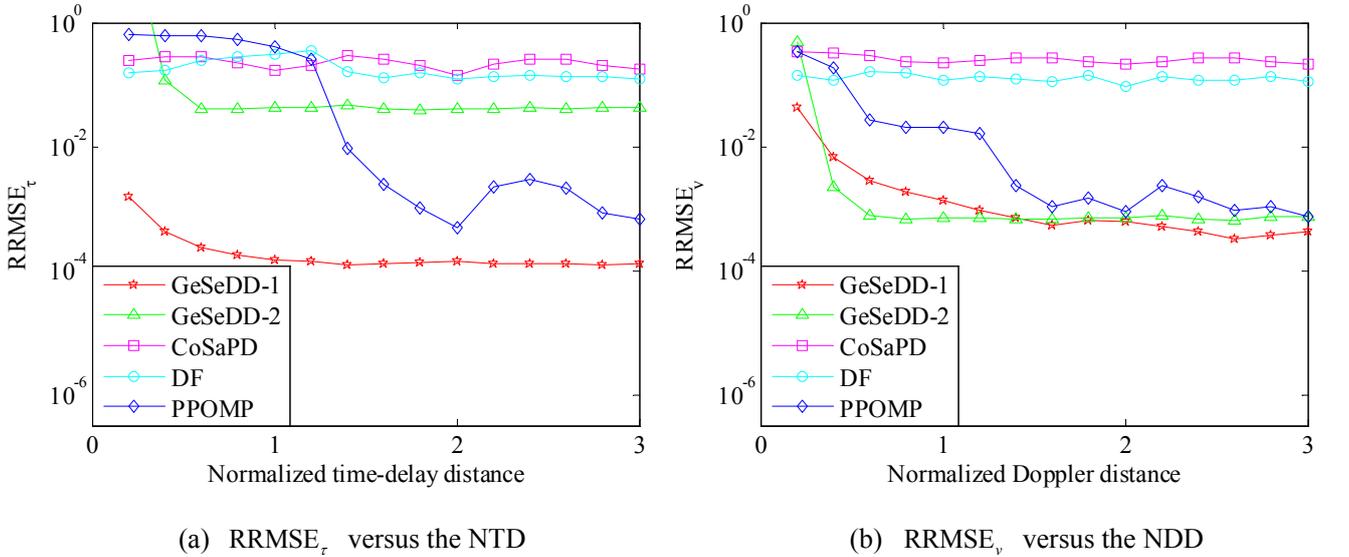

(a) $\text{RRMSE}_\tau$ versus the NTD       (b) $\text{RRMSE}_v$ versus the NDD

Fig. 2 The resolution performance with SNR=30dB

*A. Estimation Performance*

In the first experiment, we evaluate the estimation performance of all four methods in the AWGN environment. To reduce interaction between different targets, the delay separation and Doppler separation between any two targets are set to at least $2\tau_0$ and $2v_0$, respectively. Assume that there exist 10 targets. The RRMSE performance versus the SNR is shown in Fig. 1. It is seen that, when SNR $> -10$dB, GeSeDD-1 and PPOMP are superior to the other three methods. This is because both GeSeDD-1 and PPOMP are parametric estimation methods. For large SNR, GeSeDD-1 performs better than PPOMP does. This case occurs due to the numerical precision of PPOMP [8]. For the grid space setting, GeSeDD-2 has moderate estimation accuracy of the delay parameters. If we decrease the grid space, more accurate estimation can be obtained. For a small SNR (SNR $\leq -10$dB), PPOMP offers the highest estimation accuracy. The reason is that PPOMP firstly detects target locations roughly on discrete delay-Doppler grids by the orthogonal matching pursuit algorithm, which is equivalent to conduct two-dimensional matched filtering in compressive domain, and thus has robust performance in noisy case. For the proposed scheme, the beamspace MUSIC algorithms suffer from the threshold effect of SNR [32]. Then in low SNR, PPOMP is superior to the proposed scheme. In spite of the SNR, both CoSaPD and DF have poor estimation accuracy because of the basis mismatch effect. This further verifies the effectiveness of the proposed scheme for the off-grid target estimation.

*B. Resolution Performance*

In the second simulation, we test the resolution ability of two closely located targets characterized by parameters $(\tau_1, v_1)$ and $(\tau_2, v_2)$ with the same reflectivity amplitudes. Without loss of generality, assume that $\tau_1 \leq \tau_2$ and $v_1 \leq v_2$. Denote $\Delta\tau = (\tau_2 - \tau_1)/\tau_0$ and $\Delta v = (v_2 - v_1)/v_0$ as the normalized time-delay distance (NTD) and the normalized Doppler distance (NDD) of the two targets, respectively. The time delay and Doppler shift of the first target are randomly generated, while those for the second one are set to be $\tau_2 = \tau_1 + \Delta\tau\tau_0$ and $v_2 = v_1 + \Delta v v_0$. The SNR is set to 30dB. Fig. 2(a) displays the variation of the $\text{RRMSE}_\tau$ as a function of the NTD for two closed targets with the same Doppler shift. To exclude the coherent sources as required by Theorem 2, an extra target is introduced, which has the same time-delay as the second target but different Doppler shift. It is seen that GeSeDD-1 has the lowest $\text{RRMSE}_\tau$ for different NTDs and maintains a high resolution. In contrast, PPOMP cannot identify the two close targets when the NTD is small. For a large value of NTD, PPOMP can resolve the two targets but with a low estimation accuracy. On the other hand, the resolution of CoSaPD and DF is mainly determined by the grid space and thus the $RRMSE_\tau$ is independent of the NTD. Fig. 2(b) shows the $RRMSE_v$ versus NDD for the two close targets with the same time delay. Similar to Fig. 2(a), the proposed scheme again achieves the best Doppler resolution performance.

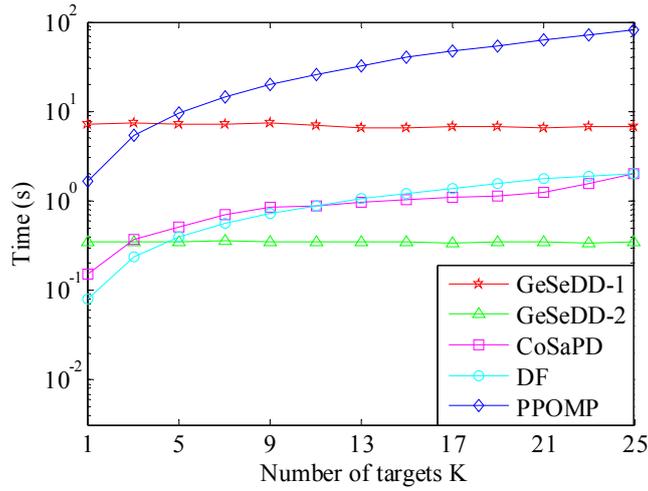

Fig. 3 The computational time versus the number of targets with SNR=20dB

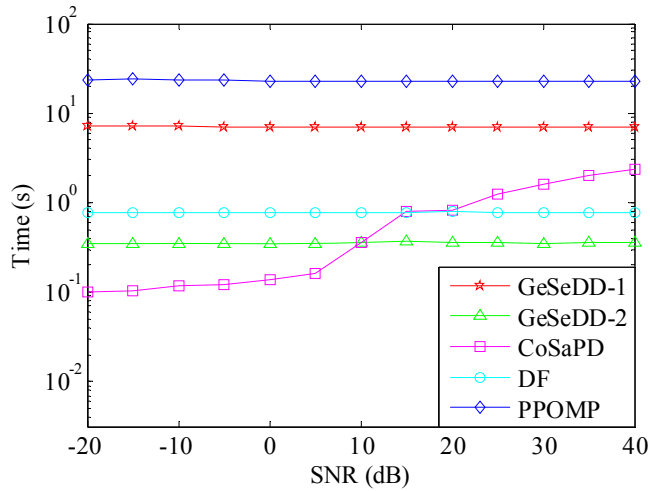

Fig. 4 The computational time versus the SNR in a noisy case ($K$=10)

## C. Computational Load

In the third simulation, we simulate the computational time of all methods. The simulation is performed in MATLAB 2011b 64-bit environment on a PC with 3.6 GHz Intel core i7-4790 processor and 16 GB RAM. Fig. 3 displays the dependence of CPU time on the number of targets $K$ with $SNR = 20 \, dB$. It is seen that GeSeDD-2 has the smallest computational cost when $K \geq 5$, while GeSeDD-1 has the largest computational cost in all sequential methods. Moreover, the consumed CPU times of both GeSeDD-1 and GeSeDD-2 almost keep unchanged for different $K$. The reason is that the CPU time of the proposed scheme is dominated by the beamspace MUSIC algorithms, which are independent on $K$. In contrary, the CPU times of DF and CoSaPD depend on the number of targets $K$. These results are consistent with the theoretical analysis in Section III-D. In Fig. 4, we show the CPU time versus the SNR with ten off-grid targets. We find that all these methods except CoSaPD have stable computational cost with different SNR. This is because the convex optimization called by CoSaPD is easy to achieve the termination condition at lower SNR.

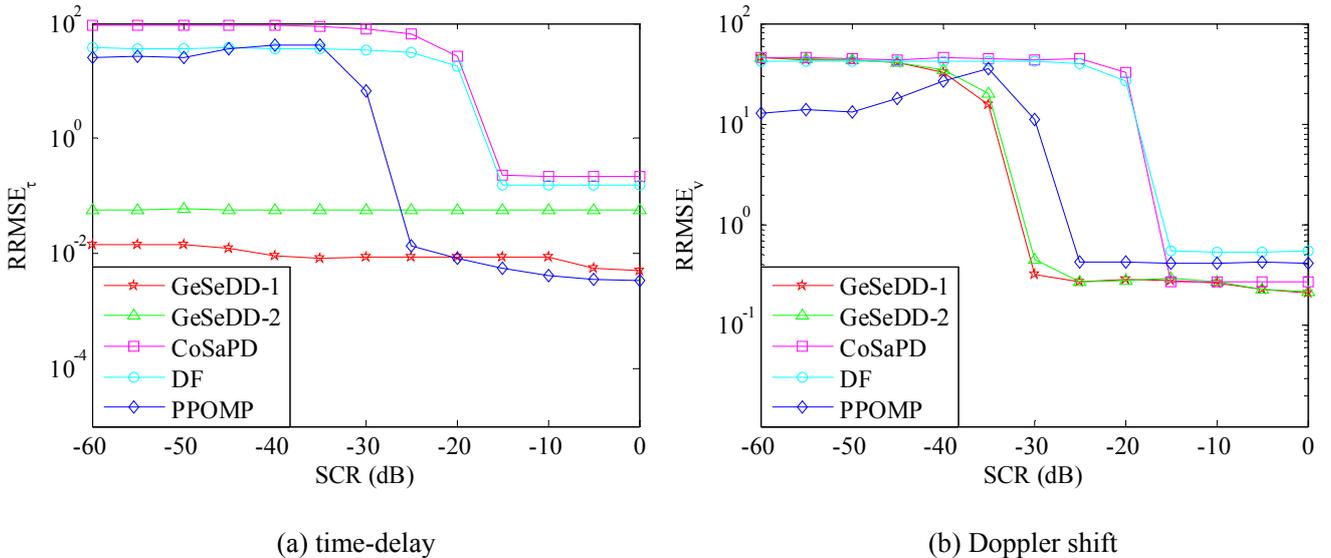

(a) time-delay  (b) Doppler shift

Fig. 5 The RRMSE versus the SCR in a noisy case ($K$=10, SNR=10dB)

*D. The Effects of Clutter*

In the fourth simulation, we demonstrate the performance of proposed scheme in the presence of clutter. It is assumed that the received echo signals are contaminated by noise and clutter simultaneously. The clutter is modeled as 4000 Swerling-0 scatterers [10], with Doppler shifts uniformly distributed in a single Nyquist bin around zero-Doppler to allow for small relative velocity. Clutter scatterers are uniformly distributed in the whole time delay interval $(0,10)\,\mu s$.

Let $\mathbf{c}_{cs}^{l} \in \mathbb{C}^{M \times 1}$ be the compressive measurements of the clutter for the $l$-th pulse. Define the signal-to-clutter ratio (SCR) as $\mathrm{SCR}_{cs}^{l} = \|\mathbf{Mr}^{l}\|_{2}^{2} / \|\mathbf{c}_{cs}^{l}\|_{2}^{2}$. As in Gaussian noise case, the SCR is assumed to be fixed in different pulse intervals. Similar to the matrix formulation in (8), the compressive measurement of the radar echo in a CPI can be expressed as

$$\mathbf{S}_c = \mathbf{M\Psi\Theta} + \mathbf{N} + \mathbf{C} \tag{24}$$

where $\mathbf{C} = \left[\mathbf{c}_{cs}^{0}, \mathbf{c}_{cs}^{1}, \cdots, \mathbf{c}_{cs}^{L-1}\right] \in \mathbb{C}^{M \times L}$ is the concatenated compressive matrix of the clutter. Note that the clutter power is mainly concentrated in the vicinity of the zero-Doppler. We can suppress the clutter by pre-processing $\mathbf{S}_c$ in Doppler domain through low-pass filtering.

In the simulations, we implement the low-pass filter with the cutoff frequency 600Hz and assume that the Doppler shifts of all 10 targets are located in the passband of the filter. Fig. 5 shows the RRMSE performance versus the SCR with SNR=10dB. It is seen that the proposed two realizations, GeSeDD-1 and GeSeDD-2, have high estimation accuracy in the presence of clutter. Especially, the proposed methods have high anti-clutter performance in the estimation of time-delays, as shown in Fig. 5(a). This is due to the parametrized model (1) and the high resolution estimates of the time delays. In performing the estimates by (24), the clutter scatters are treated as targets with different delays and small Doppler shifts. Under pre-processing $\mathbf{S}_c$ through the low-pass filtering, the clutter scatterers around the zero-Doppler shift are filtered out. Then the remaining scatterers have small amplitudes which have less effect on the delay estimation of real targets through the DOA estimation methods. In the simulated SCR range, the time delays can be well resolved and thus have high estimation accuracy. With the accurate estimation of time delays, the clutter has little effect on the noisy estimate of the coefficient matrix $\hat{\mathbf{\Theta}}$ in (14) and then the accurate estimate of the Doppler shifts is derived. In contrast, CoSaPD and DF are affected by both the clutter and the basis mismatch and they have poor performance. It is noted that PPOMP method has moderate estimation performance. This is because it does not suffer from the basis mismatch.

## VI. CONCLUSION

In this paper, we have presented a general sequential estimation scheme for time delay and Doppler shift estimation in a sub-Nyquist pulse-Doppler radar. With the parametrized model, the proposed scheme is versatile and suitable to any AIC system reported in the literature. More importantly, we theoretically derive the sufficient condition to guarantee the successful estimation of the time delays. Benefitting from the well-developed spectrum estimation techniques, the proposed scheme achieves a high estimation accuracy and a high resolution.

## ACKNOWLEDGEMENTS

The authors would like to thank the anonymous reviewers for their valuable comments. This research is supported partially by the National Natural Science Foundation of China (Nos. 61401210, 61571228 and 61671245).